\documentstyle[11pt,aaspp4,epsf]{article}
\newcommand{\simgt}{\lower.5ex\hbox{$\; \buildrel > \over \sim \;$}}
\newcommand{\simlt}{\lower.5ex\hbox{$\; \buildrel < \over \sim \;$}}

\slugcomment{NAOJ-Th-Ap 2000 No.9, KUNS-1670, HUPD-0006}
\lefthead{Liou et al.}
\righthead{Nonlinear Evolution of Very Small Scale Cosmological Baryon
Perturbations .....}

\begin{document}
\title{ Nonlinear Evolution of Very Small Scale Cosmological 
Baryon Perturbations at Recombination}

\author{Guo-Chin Liu${}^{1),2)}$, ~Kazuhiro Yamamoto${}^{3)}$,
   ~Naoshi Sugiyama${}^{1)}$, and Hiroaki Nishioka${}^{3)}$}
\bigskip
\affil{${}^{1)}$Division of Theoretical Astrophysics, National
Astronomical Observatory Japan, Mitaka, 
181-8588,~Japan}
\affil{${}^{2)}$Department of Physics, Kyoto University, Kyoto
606-8502,~Japan}
\affil{${}^{3)}$Department of Physics, Hiroshima University,
Higashi-Hiroshima 739-8526,~Japan}

\begin{abstract}

The evolution of baryon density perturbations on very small scales is
investigated.  In particular, the nonlinear growth induced by the radiation
drag force from the shear velocity field on larger scales during the
recombination epoch, which is originally proposed by Shaviv in 1998, is
studied in detail.  It is found that inclusion of the diffusion term
which Shaviv neglected in his analysis results in rather mild growth
whose growth rate is $\ll 100$ instead of enormous amplification $\sim
10^4$ of Shaviv's original claim since the diffusion suppresses the
growth.  The growth factor strongly depends on the amplitude of the
large scale velocity field.

The nonlinear growth mechanism is applied to density perturbations of
general adiabatic cold dark matter (CDM) models.  In these models, 
it has been found in the
previous works that the baryon density perturbations are not completely
erased by diffusion damping if there exists gravitational potential of
CDM.  With employing the perturbed rate equation which is derived in
this paper, the nonlinear evolution of baryon density perturbations is
investigated.  It is found that: (1) The nonlinear growth is larger for
smaller scales.  This mechanism only affects the perturbations whose
scales are smaller than $\sim 10^2M_\odot$, which are coincident with the
stellar scales.  (2) The maximum growth factors of baryon density
fluctuations for various COBE normalized CDM models are typically less
than factor $10$ for $3-\sigma$ large scale velocity peaks.  (3) 
The growth factor depends on $\Omega_{\rm b}$.

\end{abstract}



\section{INTRODUCTION}
\def\II{{\rm I\hspace{-0.5mm}I}}
\def\III{{\rm I\hspace{-0.5mm}I\hspace{-0.5mm}I}}
\def\IV{{\rm I\hspace{-0.5mm}V}}
\def\VI{{\rm V\hspace{-0.5mm}I}}
\def\VII{{\rm V\hspace{-0.5mm}I\hspace{-0.5mm}I}}
\def\rmcomv{{}}
\def\N{N}
\def\aeq{{a_{\rm eq}}}
\def\keq{{k_{\rm eq}}}
\def\Th{\Theta_{2.726}}
\def\xe{{x_e}}
\def\ne{{n_{\rm e}}}
\def\fnu{{f_\nu}}
\def\The{\Theta_{2.726K}}
\def\yp{{y_{\rm p}}}
\def\nb{{n_{\rm b}}}
\def\MJ{{M_{\rm J}}}
\def\Msolar{{M_{\odot}}}
\def\Mpc{{\rm Mpc}}
\def\calA{{\cal A}}
\def\kphys{{k^{\rm phys}}}
\def\kcomv{{k^{\rmcomv}}}
\def\rhob{{\rho_{\rm b}}}
\def\Omegab{{\Omega_{\rm b}}}
\def\Omegam{{\Omega_{\rm 0}}}
\def\Te{{T_{\rm b}}}
\def\cf{{c_{\rm f}}}
\def\cs{{c_{\rm s}}}
\def\ce{{c_{\rm e}}}
\def\me{{m_{\rm e}}}
\def\aova{{a\over a_0}}
\def\phys{{\rm phys}}
\def\deltab{{\delta_{\rm b}}}
\def\deltac{{\delta_{\rm c}}}
\def\deltam{{\delta_{\rm m}}}

Structure formation of the universe from the large scale structure to 
the stellar objects is one of the most important
issues of modern cosmology and astrophysics. 
Within a framework of the
hierarchical clustering scenario which is suggested by the adiabatic
cold dark matter (CDM) models, formation of first luminous objects in
the universe has been discussed by many authors (e.g., \cite{FK};
\cite{GO}; \cite{OG}; Haiman \& Loeb 1997; Nishi, et~al 1998).
According to those theoretical investigations, in which the initial
dark matter and baryon density fluctuations have been solved linearly, 
the first luminous objects are thought to appear in the gas clouds 
in the dark halo
at $z\sim{{\rm a~few}\times 10}$, depending on the cosmological
models.

On the other hand a new mechanism which causes rapid growth of the
baryon density fluctuations on very small scales during a short period
of the recombination epoch has been proposed by Shaviv (1998).  This
growth of the baryon density perturbations is caused by 
the radiation drag force from the large scale velocity field  
which is coupled to the small scale baryon
perturbations in a nonlinear manner.  He has pointed out enormous
growth ($\sim 10^4$) of baryon fluctuations if the wavelength of the
perturbations is shorter than a critical value.  Such enormous
amplification of baryon density fluctuations must lead to early
formation of stellar objects.  If the amplification is as large as
Shaviv's prediction, therefore, it might be a quite interesting
antithesis for the standard scenario of the cosmic structure formation
in the CDM cosmological models, though we will show that this nonlinear
growth mechanism would not be so drastic.

In this paper, we first carefully re-examine Shaviv's treatment of 
density perturbations.  We find the diffusion term which Shaviv
has neglected in his analysis plays a crucial role for suppression 
of the nonlinear growth.   

There are also some artificial assumptions in Shaviv's investigation.
First one is in use of the thermal equilibrium for the recombination
process.  This nonlinear growth is induced by 
the shear velocity field on larger scales through the radiation drag
force against density fluctuations of ionized hydrogens.
Therefore we need to calculate fluctuations of the ionization
fraction.  Shaviv has obtained these fluctuations 
by employing the perturbed Saha equation which should not be applied 
during the recombination epoch.   Here we develop 
a new method to treat these fluctuations by the perturbed treatment of
the rate equation which is 
an extension of the previous work by Yamamoto,
Sugiyama, and Sato (1997, Paper I).

The next assumption Shaviv has made in his investigation is the
existence of isothermal baryon density fluctuations.  
It is well known that the
adiabatic baryon and photon density perturbations are erased away due
to the photon diffusion before and during the recombination epoch
(Silk 1968;~Sato 1971;~Weinberg 1971).
The diffusion (Silk) damping scale goes over the galaxy scale.
Therefore it seems to be unavoidable for the nonlinear growth mechanism
to assume small scale isothermal baryon
perturbations during the recombination epoch as the seeds
of baryon perturbations in an artificial manner.
The origin of the small scale isothermal baryon perturbations has
not been considered.   

However we would like to stress that the small scale isothermal baryon
density fluctuations are produced after the Silk damping in general CDM
cosmological models without any ad hoc assumptions.  Such isothermal
baryon density fluctuations on small scales have been found by Yamamoto,
Sugiyama \& Sato (1998, Paper II).  Once baryon fluctuations are damped
by diffusion.  However they grow again owing to 
the gravitational potential of
CDM before the recombination epoch due to the breakdown of the
tight-coupling between baryons and background photons.  Since the
Compton drag from background photons is still effective at that epoch
and works as  friction, this growth of baryon density fluctuations
is characterized by the terminal
velocity.  Such natural existence of the small scale isothermal baryon
density fluctuations motivates us to revisit Shaviv's investigation in the
context of the conventional adiabatic CDM models.

The paper is organized as follows: In \S 2, we review the nonlinear
growth mechanism which Shaviv has proposed.  Some numerical tests for a
toy model with simple assumptions are presented in \S 3.  
It is shown that the diffusion term plays a crucial role for the
estimate of the growth factor.  
In \S 4, we examine
whether the nonlinear growth mechanism is effective on the 
adiabatic CDM models or not.
\S 5 is devoted to summary. Throughout this paper
we work in units where $c=\hbar=k_B=1$, and assume
$T_0=2.726~{\rm K}$ as the cosmic microwave radiation
temperature at present.

\section{Basic Equations for the Nonlinear Effect}

Before describing the nonlinear growth mechanism of small scale 
baryon density fluctuations proposed by Shaviv (1998),
let us first summarize basic 
equations (see also Paper I; II). We work in the Newtonian perturbed metric,
\begin{equation}
  ds^2=\biggl({a\over a_0}\biggr)^2
  \Bigl(-(1+2\Psi)d\eta^2+(1-2\Psi)d{\bf x}^2\Bigr),
\end{equation}
where $\Psi$ is the perturbed gravitational potential,
$a$ is the scale factor whose suffix $0$ indicates the
present value, and $\eta \equiv \int dt (a_0 /a)$. 
As we are interested in the 
small scale
cosmological perturbations, we can assume the geometry
of the universe to be flat.
Then, in the high redshift universe,
the Friedmann equation leads to
\begin{equation}
  H^2=\biggl({\dot a\over a }{a_0\over a}\biggr)^2=
  \biggl({a_0\over a}\biggr)^4 {\aeq+a\over \aeq+a_0} \Omega_0 H_0^2,
\end{equation}
where the dot denotes $\eta$-differentiation,
$H_0$ is the Hubble constant, and $\aeq$ is the
scale factor at the matter-radiation equality epoch
which can be written as
\begin{equation}
{a_0\over \aeq}=4.04\times10^{4}(1-f_\nu)
  \Omega_0h^2,
\end{equation}
with $f_\nu$ being the neutrino fraction in the energy density of
radiation components. 
For the massless standard neutrino model with three species,
$f_\nu\simeq 0.405$, which we adopt throughout the present paper.
Here, we take into account 
the fraction of helium abundance.  Hence 
we define the fractional ionization $x_e$ as
\begin{equation}
x_e \equiv \frac{n_e}{2n_{\rm He}+n_{\rm H}} ,
\end{equation}
where $n_e$, $n_{\rm He}$, and $n_{\rm H}$ are the number densities of free 
electron, helium, and hydrogen, respectively. 
The primordial helium mass fraction is defined as 
\begin{equation}
y_p \equiv \frac{4n_{\rm He}}{n_{\rm b}},
\end{equation}
where $n_{\rm b} \equiv 4n_{\rm He} + n_{\rm H}$ 
is the net baryon number density. We set $y_p=0.24$ throughout 
the 
present paper. 
The fractional ionization $x_e$ is $1$
in the very early universe and becomes $(1-y_p)/(1-y_p/2) \simeq 0.86$
just before the 
recombination epoch due to the early recombination of 
helium atoms. 

Let us now describe the nonlinear growth mechanism.
We proceed with reviewing equations of the small
perturbations for the baryon-electron fluid. Since we are
interested in the evolution of density perturbations on very
small scales, we adopt the Newtonian approximation.
The evolution equations for the baryon-electron fluid
coupled with the primeval radiation through Compton scattering
are written as
\begin{eqnarray}
  &&{\dot\rhob}
  +3{\dot a\over a}\rhob +\nabla_i(\rhob V_{\rm b}^{i})=0,
\label{dotdeltab}
\\
  &&\dot V_{\rm b}^i+{\dot a\over a}V_{\rm b}^i+V_{\rm b}^j \nabla_j 
V_{\rm b}^i+
  {1\over \rhob}\nabla^i P+ \nabla^i \Psi
  ={a\over a_0}{\ne \sigma_T\over R}(V_\gamma^i-V_{\rm b}^i),
\label{Vbeq}
\end{eqnarray}
where $\rhob$ and $V_{\rm b}^i$ are the density and the velocity
of the baryon-electron fluid, respectively, $P$ is 
the pressure, $\sigma_T$ is the Thomson cross section, $V_\gamma^i$
is the dipole moment of the photon field, and $R$ is defined
as $R \equiv {3\rhob/ 4\bar{\rho}_\gamma}$ with the
spatially averaged energy density of the photon field
$\bar{\rho}_\gamma$, 
where $\bar{\null~\null}$ denotes the background quantity. 
The right hand side of equation (\ref{Vbeq}) represents the radiation
drag force caused through the Compton interaction between electrons and
photons which can be written by using $\tau_{\rm D}$ that is the time 
scale of momentum transfer between the baryon-electron fluid and the 
photon fluid as
\begin{equation}
  {a\over a_0}{4 \sigma_T\bar{\rho}_\gamma x_{e}\over 3m_p}
  \left(1-{y_p \over 2}\right)
  \left(V_\gamma^i-V_{\rm b}^i\right) \equiv {1\over\tau_{\rm D}}({V_\gamma^i-V_{\rm b}^i}),
\label{radiationforce}
\end{equation}
where $m_p$ denotes the proton mass.
As the universe expands, the momentum transfer time
scale $\tau_{\rm D}$ becomes longer.
However the drag time scale is still small enough to keep 
the tight coupling between baryon-electron and photon fluids
even just before the recombination epoch, 
so that $V_\gamma^i=V_{\rm b}^i$.
Eventually the tight coupling
breaks down as the recombination process proceeds when 
the photon mean free path becomes larger
than the wavelength of the density perturbations (see also Paper II).

Let us now focus on perturbations of the baryon-electron fluid on very 
small scales.  
As far as considering linear density perturbations,  
perturbations of each scale evolve individually.
Following Shaviv (1998), however, we take into account 
the quasi-nonlinear contribution of the drag force from   
the large scale velocity field.
To be specific, we separate the velocity field into
the small scale part which we focus on and 
the large scale part which describes the nonlinear contribution
to the small scale perturbations as,
\begin{equation}
  V_\gamma^i-V_{\rm b}^i=V_{\gamma{(\rm S)}}^i-V_{\rm b{(\rm S)}}^i
                  +V_{\gamma{(\rm L)}}^i-V_{\rm b{(\rm L)}}^i ,
\end{equation}
where indices ${(\rm S)}$ and ${(\rm L)}$ indicate small and large scales,
respectively.
The fluctuation of the fractional ionization on the small scale
is defined as $\delta x_{e{(\rm S)}} \equiv x_e-\bar{x}_e$. 
Then the
right hand side of equation (\ref{Vbeq}) leads to 
\begin{equation}
  {a\over a_0}{4 \sigma_T\bar\rho_\gamma \over 3 m_p}
\left(1- {y_p \over 2}\right)
  \Bigl(\bar x_{e}(V_{\gamma{(\rm S)}}^i-V_{\rm b{(\rm S)}}^i)
  +\delta x_{e{(\rm S)}} \Delta V_0^i\Bigr),
\label{rhs}
\end{equation}
where 
\begin{equation}
  \Delta V_0^i \equiv V_{\gamma{(\rm L)}}^i-V_{\rm b{(\rm L)}}^i .
\end{equation}
Here we ignore higher order terms of the small scale perturbations.
The term $\delta x_{e{(\rm S)}} \Delta V_0^i$ in equation (\ref{rhs})
describes the quasi-nonlinear coupling
between the large scale shear velocity and the small scale baryon
density fluctuations.  
Adding this quasi-nonlinear term, we can write
the linear perturbation equations on the small scale as
\begin{eqnarray}
  &&\dot\delta_{\rm b{(\rm S)}}+\nabla_iV_{\rm b{(\rm S)}}^i=0,
\label{contE}
\\
  &&\dot V_{\rm b{(\rm S)}}^i+{\dot a\over a} V_{\rm b{(\rm S)}}^i+c_s^2
  \nabla_i\delta_{\rm b{(\rm S)}}
  +\nabla_i\Psi_{{(\rm S)}}
  ={a\over a_0}{4 \sigma_T\bar\rho_\gamma \over 3 m_p}
  \Bigl(-\bar x_eV^i_{\rm b{(\rm S)}}+\delta x_{e{(\rm S)}}\Delta V_0^i\Bigr),
\label{ShavivVb}
\end{eqnarray}
where $\cs$ is the sound velocity of the baryon-electron fluid
defined by $\cs^2={\dot P}/{\dot \rhob}$. 
In deriving equation (\ref{ShavivVb}),
we have assumed the isothermal perturbations for
the small scale perturbations, i.e., $V^i_{\gamma(\rm S)}=0$
since the Silk damping process erases
the density perturbation and the dipole moment
of the photon field on small scales.

Let us investigate the effect of the quasi-nonlinear term 
$\delta x_{e{(\rm S)}}\Delta V_0^i$ on 
the evolution of perturbations. 
Before the recombination epoch, $\Delta V_0^i=0$ due to the tight 
coupling. However, the
tight coupling becomes broken down, i.e., $\Delta V_0^i\neq0$,
as the decoupling process proceeds and the contribution from 
the quasi-nonlinear term may not be negligible anymore. 
In fact Shaviv has claimed that 
this quasi-nonlinear contribution, which represents the
spatially fluctuating radiation drag force, causes enormous 
amplification of
baryon density fluctuations during the recombination epoch. 
Hereafter, we omit the label '{(\rm S)}', for simplicity. Combining 
equations (\ref{contE}) and (\ref{ShavivVb}),
we obtain 
\begin{equation}
  \ddot\deltab+{\dot a\over a}\dot\deltab-\cs^2\triangle\deltab
  -\triangle\Psi=-\frac{1}{\tau_{\rm D}}\deltab
  \Bigl(1+\frac{\Delta V_0^i \nabla_i\delta x_{e}}{\bar x_e\deltab}\Bigr) ,
\end{equation}
where $\triangle$ is the Laplacian.  

By assuming the wave form solution,
$\deltab\propto e^{(\omega\eta-i\bf x \cdot\bf k)}$,
where $\bf k$ is the comoving wave number, we have the dispersion relation
\begin{equation}
  \omega^2+\biggl({1\over \tau_{\rm H}}+{1\over \tau_{\rm D}}\biggr)\omega
  +\biggl({1\over\tau_{\rm o}^2}-{1\over\tau_{\rm J}^2}-i{1\over\tau_{\rm R}^2}\biggr)
  =0~ ,
\label{dispersion}
\end{equation}
where
\begin{eqnarray}
  &&{1\over\tau_{\rm H}}={\dot a\over a}=3.2\times10^{-18} \sqrt{\Omega_0 h^2}
  \sqrt{1+z}~{\rm s}^{-1} ,
\label{tauH}
\\
  &&{1\over \tau_{\rm D}}={a\over a_0}
    {4 \sigma_T\bar\rho_\gamma \bar x_{e}\over 3 m_p} 
\left(1-{y_p \over 2}\right)
  =7.4\times10^{-24} \bar x_e (1+z)^3~{\rm s}^{-1} ,
\label{deftauD}
\\
  &&{1\over\tau_{\rm o}}=\cs k=1.61\times10^{-16}
  (\Omegab h^2)^{1/3}\biggl({M\over \Msolar}\biggr)^{-1/3}
  \sqrt{1+\bar x_e}\sqrt{1+z}~{\rm s}^{-1}  ,
\\
  &&{1\over\tau_{\rm J}}=\sqrt{4\pi G \Bigl({a\over a_0}\Bigr)^2\rhob}
  =\sqrt{3\Omegab\over 2\Omega_0}{1\over \tau_{\rm H}} ,
\\
  &&{1\over\tau_{\rm R}}=\sqrt{|\alpha| \Delta V_0 k
  \over \tau_{\rm D}}
  =4.9\times10^{-17}(\bar x_e|\alpha| \Delta V_0)^{1/2}
  (\Omegab h^2)^{1/6} (1+z)^{3/2} \biggl({M\over \Msolar}\biggr)^{-1/6}
  ~{\rm s}^{-1} ,
\label{tauR}
\end{eqnarray}
with $k\equiv \vert{\bf k}\vert$ and $\Delta V_0\equiv \Delta V_0^i
\cdot k^i/k$ being the shear velocity from the large scale.  Here
$\tau_{\rm H}$, $\tau_{\rm D}$, $\tau_{\rm o}$, $\tau_{\rm J}$, and $\tau_{\rm R}$
are time scales for the Hubble expansion, the radiation drag, the sound
oscillation, the Jeans oscillation, and the quasi-nonlinear radiation
force, respectively. The parameter $\alpha$ of equation ({\ref{tauR}})
is defined as
\begin{equation}
\alpha \equiv \frac{\delta x_e}{\bar x_e\deltab} .
\label{alpha}
\end{equation}
A nonzero value of $\alpha$ provides the quasi-nonlinear effect.
Shaviv obtained $\alpha$ by employing the Saha equation. 
The validity of this assumption will discuss in \S4.
The derivation of $\delta x_e$ is shown in Appendix B.
In the above expressions, the baryon mass $M$ is used instead of
the comoving wave number $k$ (Paper I), which is defined by
\begin{equation}
  M={4\pi \bar{\rho}_{\rm b} \over 3}\biggl({\pi\over k}{a\over a_0}\biggr)^3 .
\label{bmass}
\end{equation}
It should be noticed that we take into account 
only the baryon component for the
gravitational potential.

Let us now evaluate the growth rate of the growing-mode solution.
The solution of the dispersion relation (\ref{dispersion})
can be written as 
\begin{equation}
  \omega={1\over 2}\Biggl[ -\biggl({1\over \tau_{\rm H}}+{1\over \tau_{\rm
D}}\biggr)
  +\sqrt{\biggl({1\over \tau_{\rm H}}+{1\over \tau_{\rm D}}\biggr)^2
  -4\biggl({1\over\tau_{\rm o}^2}-{1\over\tau_{\rm J}^2}-i{1\over\tau_{\rm
R}^2}\biggr)}
\Biggr]~.
\label{solomega}
\end{equation}
Here we omit the negative solution which provides the decaying mode 
solution.
The positive real part of $\omega$, i.e, $\Re[\omega(\eta)]$, 
represents the growth rate of the
instability.  
The exact expression of $\Re[\omega(\eta)]$ is shown in
Appendix A. 
In the time interval between $\eta_1$ and $\eta_2$, 
the amplification of the baryon density fluctuations, which 
we refer the growth factor $D(\eta_2, \eta_1)$, 
becomes
\begin{equation}
D(\eta_2, \eta_1) \equiv \frac{\deltab(\eta_2)}{\deltab(\eta_1)}=
\exp\Bigl(\int_{\eta_1}^{\eta_2} d\eta {\Re}[\omega(\eta)]\Bigr) .
\label{growthfactor}
\end{equation}

\section{Nonlinear Growth of a Toy Model}
Before carrying out fully consistent calculations based 
on density perturbations of actual cosmological models, 
we investigate the nonlinear growth 
for a toy model.  First, we assume the large scale velocity 
$\Delta V_0$ is constant in time throughout this section.   
Secondly, as Shaviv has done in his 
paper, the fractional ionization $\bar{x}_e$ and 
its perturbation $\delta x_e$ are calculated by employing the Saha 
equation, which is described in Appendix B.  
We take the fiducial CDM model, i.e.,  
$\Omega_0=1.0,\Omegab=0.04,$ and  $h=0.5$.
The time evolution of the fractional ionization is shown in 
Fig.~1.  
\begin{figure}
\centerline{\epsfxsize=15cm \epsffile{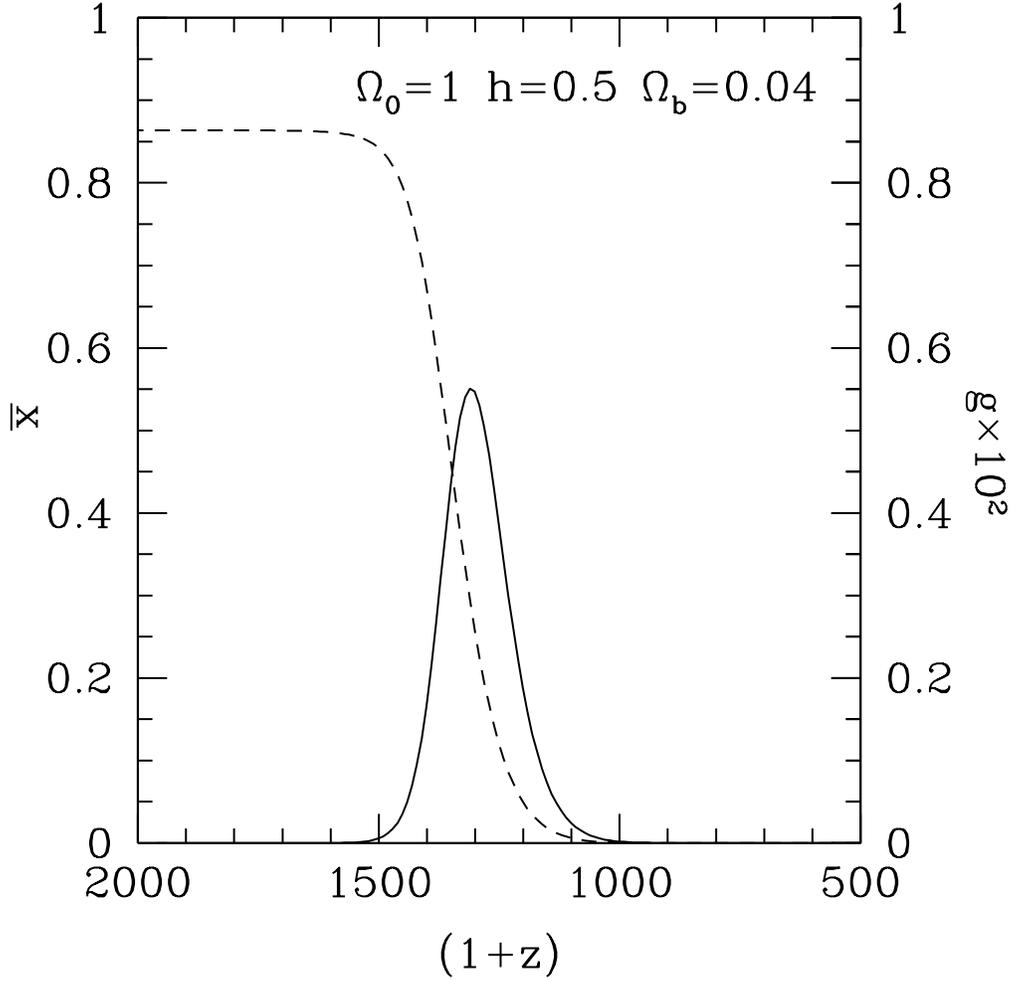}}

\caption{Fractional ionization $\bar{x}_e$ (dashed line) 
and visibility function
 $g (z)$ (solid line) from the Saha equation with the CDM model; 
$\Omega_0 = 1.0,~\Omegab = 0.04$, and $h = 0.5$. The peak of the 
visibility function ($z=1300$) corresponds
 to the recombination epoch. In this fiducial CDM model, the recombination
occurs in between   $1+z = 1500$ and $1100$.
}
\label{Fig. 1}
\end{figure}
We also plot the redshift visibility function 
\begin{equation}
g (z) \equiv - {d\tau \over dz} \rm{e}^{-\tau}, 
\end{equation}
where 
$\tau \equiv -\int_{\displaystyle \eta}^{\displaystyle\eta_0} d\eta' 
\displaystyle{a \over a_0}n_e\sigma_T $ is the optical depth.
The width of the visibility function corresponds to the recombination epoch.
Therefore the recombination epoch of this model is $1+z = 1500-1100$.

Let us first compare time scales of equations 
(\ref{tauH})--(\ref{tauR}).
In Fig.~2, these (inverse) time scales are plotted as a function of
$(1+z)$.  
\begin{figure}
\centerline{\epsfxsize=15cm \epsffile{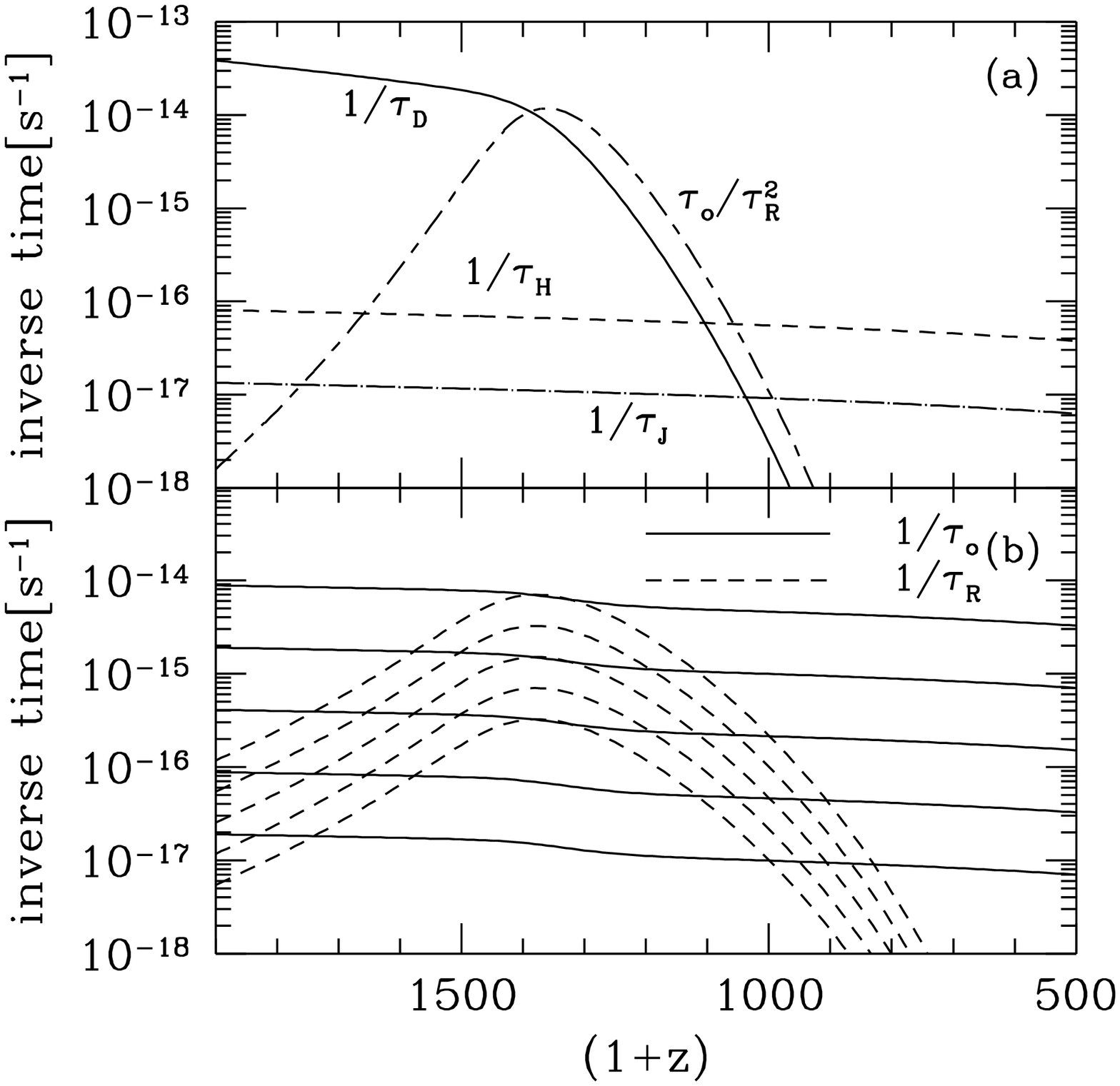}} 
\caption{
Inverse time
scales (eq.~[\protect\ref{tauH}] - eq.~[\protect\ref{tauR}]) 
for the fiducial CDM model as in Fig.~1.  
The Saha equation is employed to calculate the fractional ionization and 
$\Delta V_0 =10^{-4}$ for $\tau_{\rm R}$.
Wave number independent and dependent time scales are plotted
 in (a) and (b), respectively.
In panel (a), $\tau_{\rm o}/ \tau_{\rm R}^2$ which provides the 
maximum growth rate together with $\tau_{\rm D}$ 
(eq.[\protect\ref{max}]) is also plotted
by a long dashed-short dashed line.  
In panel (b), the wave numbers of the baryon density perturbations
are $k=71.2, 330, 1530, 7120$, and $33000{\rm Mpc}^{-1}$ 
which correspond to mass scales $10^{6}, 10^{4}, 10^{2},
10^{0}$, and $10^{-2} M_{\odot}$, respectively, from bottom to top.  
} 
\label{Fig. 2}
\end{figure}
The panel (a) shows $1/\tau_{\rm D}$, $1/\tau_{\rm H}$, and $1/\tau_{\rm 
J}$,
all which do not depend on $k$.  
Note that $1/\tau_{\rm D}$ dominates others before and during
the recombination epoch.  
However as the recombination process proceeds,
this time scale suddenly increases
and eventually exceeds other time scales after recombination.
In this panel, we plot $\tau_{\rm o}/\tau_{\rm R}^2$ which 
is also independent on $k$.  We will discuss this variable later in this 
section. 

In the panel (b), 
we plot $1/\tau_{\rm o}$ and $1/\tau_{\rm R}$ for
different wave numbers $k$, i.e.,  
$k= 71.2$, $330$, $1530, 7120$, and $33000{\rm Mpc^{-1}}$
which correspond to mass scales $10^{6}, 10^{4}, 10^{2}, 10^{0}$, and $10^{-2} 
M_{\odot}$, respectively.

Now let us discuss the growth rate $\Re[\omega(\eta)]$ by comparing the
time scales in the expression (\ref{solomega}).  As is shown in Fig.~2,
the larger the wave number $k$, the larger the inverse time scales
$1/\tau_{\rm R}$ and $1/\tau_{\rm o}$ become.  Therefore we expect to
obtain the maximum growth rate when we take the small scale limit, i.e.,
$k \rightarrow \infty$.  As is shown in Appendix A, the real part of the
square root term of equation (\ref{solomega}) approaches to $\tau_{\rm
o}/\tau_{\rm R}^2$ which is the combination seen in the panel (a) of
Fig.~2.  Shaviv claimed that this combination provides the maximum
growth rate, i.e.,
\begin{equation}
\Re[\omega(\eta)]_{\rm max}^{\rm Shav} = {\tau_{\rm o}\over 2 \tau_{\rm R}^2 }
= {\alpha \Delta V_0 \over 2 c_s \tau_{\rm D}} ~ . 
\label{max_s}
\end{equation}

However, this is not the whole story.  
The complete expression of 
equation (\ref{solomega}) with large $k$ limit becomes
\begin{equation}
\Re[\omega(\eta)]_{\rm max}\equiv \lim_{k\rightarrow \infty}~
\Re[\omega] = -\frac{1}{2} \frac{1}{\tau_{\rm D}} +
{\tau_{\rm o}\over 2\tau_{\rm R}^2}
  =-\frac{1}{2} \frac{1}{\tau_{\rm D}} +
{\alpha \Delta V_0 \over 2 c_s \tau_{\rm D}} ~ .
\label{max}
\end{equation}
Here we ignore the inverse Hubble time scale because it is much smaller than 
$1/\tau_{\rm D}$ during recombination.  
Remember that $1/\tau_{\rm D}$ is comparable 
to ${\tau_{\rm o}  /\tau_{\rm R}^2}$ during recombination 
as is shown in Fig.~2(a).
In fact, the existence of this $1/\tau_{\rm D}$  term suppresses the
nonlinear growth a lot.  
    From equation (\ref{max}), it is found that a
 sufficient condition for the nonlinear 
growth at least on the smallest scale is 
$\Re[\omega(\eta)]_{\rm max}>0$, i.e.,
\begin{equation}
 \alpha \Delta V_0 >c_s .
\end{equation}

In Fig.~3, the growth rates and the growth factors of the toy model 
are plotted for various wave numbers and the large wave number limit.  
\begin{figure}
\centerline{\epsfxsize=15cm \epsffile{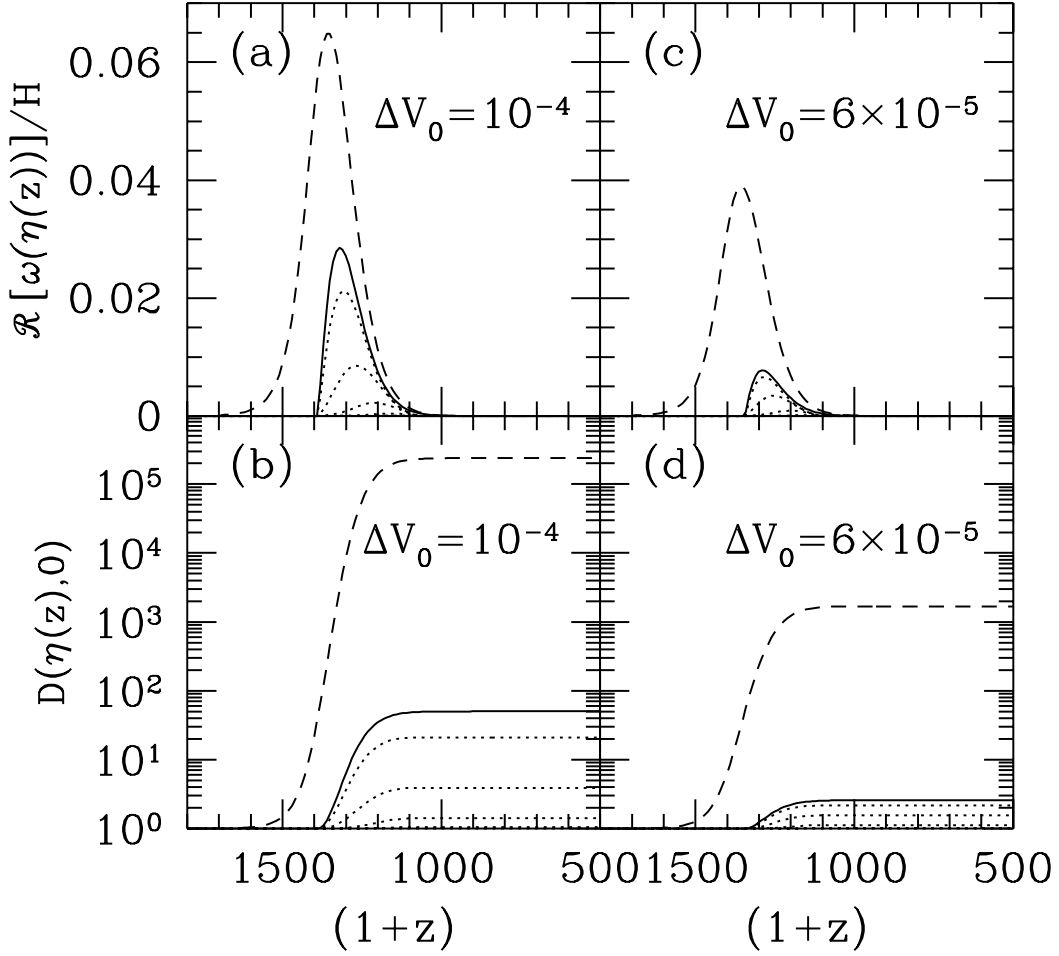}}
\caption{Growth rates $\Re[\omega(\eta(z))]$ ((a) and (c)) in the unit
of the Hubble time and growth 
factors 
$D(\eta (z), 0) \equiv 
\exp\left(\int_{0}^{\eta} d\eta' 
{\Re}[\omega(\eta')]\right) 
= 
\exp\left(\int_{1+z}^{\infty} d(1+z') 
{\Re}[\omega]/H\right) 
$  ((b) and (d)) of two 
simple simulations with fixed shear velocities 
$\Delta V_0=10^{-4}$ and $6 \times 10^{-5}$
for the same model as in Fig.~1.
The dotted lines 
plot the growth rates or growth factors for the wave numbers $k=33000, 
7120, 1530, 330$, and $71.2{\rm Mpc}^{-1}$ from top to bottom. 
One can see almost no growth for $k=330$ and $71.2{\rm Mpc}^{-1}$.
The solid lines are the maximum growth rates (eq. 
[\protect\ref{max}]) and growth factors.
 The maximum growth rates and growth factors of the investigation by Shaviv 
(eq. [\protect\ref{max_s}]) are also plotted by 
dashed lines for comparison. 
It is shown that inclusion of the diffusion term $1/\tau_{\rm D}$ (eq. 
[\protect\ref{max}]) results in the significant 
suppression of the growth factors. 
}

\label{Fig. 3}
\end{figure}

   From this figure, it is found that the growth rate and the growth factor
are very sensitive to the value of the large scale velocity field 
$\Delta V_0$.   The growth factor of $\Delta V_0 = 10^{-4}$ 
is more than factor $20$ larger than the one of $\Delta V_0 = 6\times
10^{-5}$ for the large $k$ limit. 

The ``maximum'' growth rates and growth factors obtained by Shaviv's formula
(eq.~[\ref{max_s}]) are also plotted for comparison in Fig.~3. 
It is clear that 
neglect of the diffusion results in the extremely large false growth
factor.  By the correct treatment, rather mild growth is found.  
When we take relatively large velocity $\Delta V_0 = 10^{-4}$,
we obtain about factor $50$ growth factor while
Shaviv's treatment induces more than $10^5$ growth of 
small scale density perturbations.
  From Fig.~3, it is shown that the nonlinear growth mechanism 
starts to work in the relatively early epoch ($z \sim 1500$) 
if we neglect the diffusion term.  
However, by the correct treatment, it is found that the nonlinear growth 
works only when $\tau_{\rm o}/\tau_{\rm R}^2 > 1/\tau_{\rm D}$, i.e., 
$1+z < 1340$ in this fiducial CDM model 
as is shown in Fig.~2.  Even in this shorter period, 
the maximum growth rate is suppressed by the existence of 
the diffusion term (eq.~[\ref{max}]).
We should conclude that the enormous growth $\sim 10^4$
of the baryon density fluctuations which Shaviv claimed in 
his paper is not likely to happen.

However, we should emphasize that this nonlinear growth mechanism still 
works and enhances the baryon density perturbations.  
It is interesting to estimate the nonlinear growth factor 
for realistic cosmological models.

\section{Nonlinear Growth in adiabatic CDM Models}

Since it is generally believed that adiabatic small scale baryon density 
perturbations decay
out before and during the recombination epoch due to the diffusion (Silk)
damping, an additional assumption to standard adiabatic perturbations 
seems to be needed for the nonlinear growth mechanism
which requires the existence of small scale baryon perturbations.
In the paper by Shaviv (1998), for example, he artificially 
assumed the existence of baryon
isothermal density perturbations.

However, we should point out that small scale {\it isothermal} 
baryon density perturbations 
are in fact existing at the recombination epoch
in {\it general adiabatic} CDM models.
Recently, the authors of the present paper (K.Y. \& N.S.) and Sato have
found that small scale baryon perturbations grow after
the diffusion (Silk) damping even before the recombination epoch (Paper II).
This growth of baryon density perturbations before
the recombination epoch is the result of the breaking of the
tight-coupling between the baryon perturbations and the
photon perturbations on small scales. The growth is
characterized by the terminal velocity, which implies the
balance between the gravitational force due to the potential
of CDM and the friction force by the Compton interaction with
background photons.
According to the previous investigation in Paper II,
the Fourier coefficient of the small scale baryon perturbations
grows in proportion to $(1+z)^{-7/2}$.
The small scale baryon density perturbations have amplitude of order
${\cal O}(10^{-2})\sim{\cal O}(10^{-3})$ compared with the CDM
perturbations.

The small scale photon perturbations are smeared out by the diffusion
damping.  On the other hand, the energy transfer between the
baryon-electron fluid and the background photons is effective during and
for some time after the recombination epoch.  Therefore the nature of
the small scale baryon perturbations is isothermal during these epochs
(see Paper I).
These isothermal baryon perturbations could be enhanced 
by the nonlinear growth  mechanism although the growth factor may not be 
as large as originally thought by Shaviv as is described in 
the previous section. 

The amplitude of the
large scale velocity field $\Delta V_0$ is crucial for the
instability as is shown in Fig.~3. In linear theory 
of density perturbations, the
variance of square of the large scale velocity field is evaluated
by \begin{equation}
  \bigl<\bigl| V_\gamma^i-V_{\rm b}^i\bigr|{}^2\bigr>
  ={1\over 2\pi^2}\int dk k^2
  \bigl|V_\gamma (k,\eta)^i-V_{\rm b}(k,\eta)^i\bigr|^2,
\label{shear}
\end{equation}
where $V_\gamma(k,\eta)^i$ and $V_{\rm b}(k,\eta)^i$ are
Fourier coefficients of the dipole anisotropy of
the photon field and the peculiar velocity field of
the baryon fluid, respectively.
Introducing the parameter $\nu$, we derive the
large scale velocity field from the rms one as
\begin{equation}
  \Delta V_0=\nu \sqrt{\bigl<\bigl| V_\gamma^i-V_{\rm b}^i\bigr|{}^2\bigr>}~.
\label{defnu}
\end{equation}

We consistently solve the linear perturbations
of the photon, baryon, and dark matter system in the expanding universe
to calculate the rms velocity 
on large scales.  
In Fig.~4,  the rms velocities of 
various COBE normalized CDM models are shown.  
It is found that the rms velocities are typically 
$10^{-6}$  just before recombination. They rapidly increase toward 
$10^{-4}$ as the recombination process proceeds when 
the tight coupling between photons and baryons breaks.
We may conclude that effective values of the rms velocities during
the recombination epoch are about a few times $10^{-5}$
regardless of model parameters.  
According to our toy model calculations as is shown in Fig.~3, therefore, 
we expect only high velocity peaks induce the nonlinear 
growth on small scales in the realistic cosmological models.
\begin{figure}
\centerline{\epsfxsize=15cm \epsffile{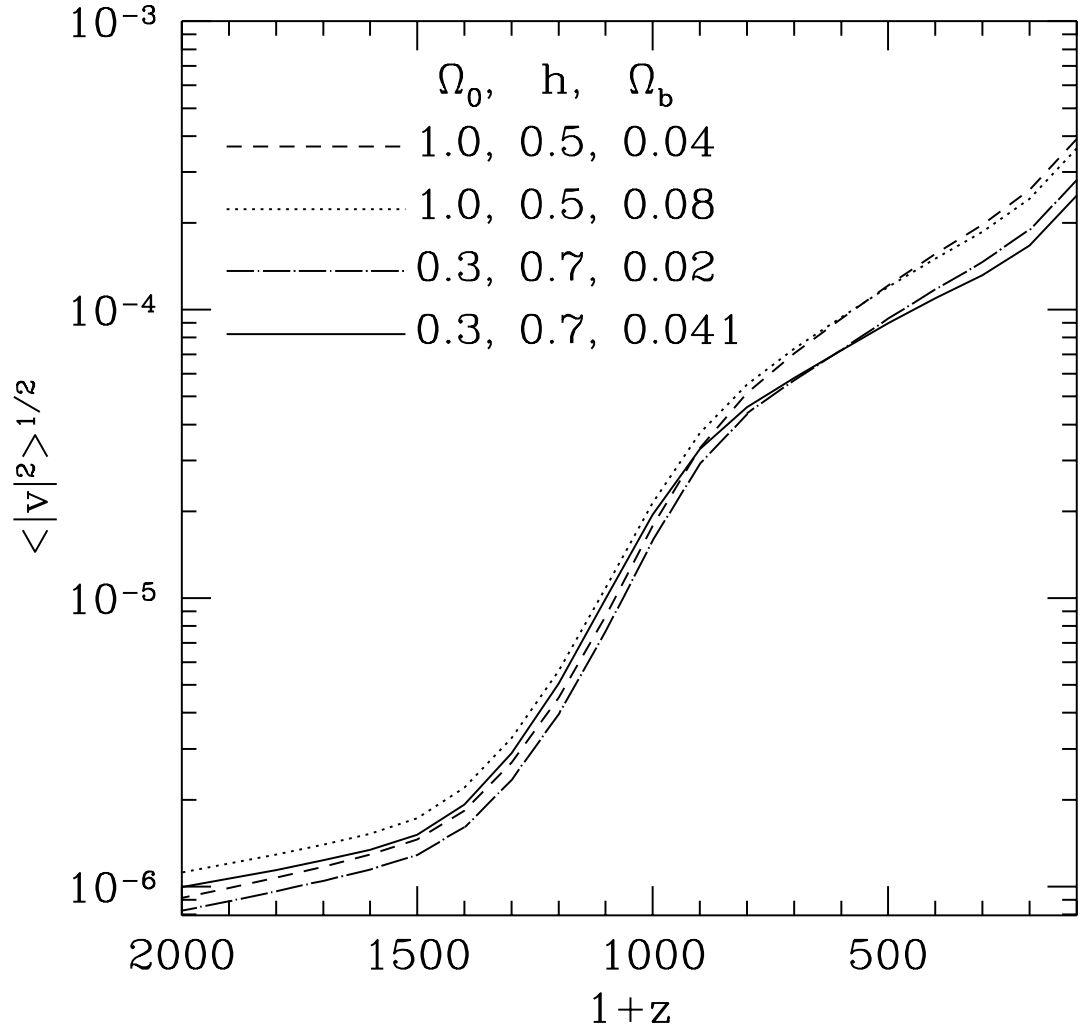}}
\caption{Root mean square shear velocities in large scales. The COBE 
 normalized velocities 
 of various flat CDM models grow
 typically from $10^{-6}$ to $7 \times 10^{-5}$ during the recombination
 epoch.
 }

\label{Fig. 4}
\end{figure}

To calculate the growth rate and the growth factor of baryon density 
fluctuations,
we need to obtain perturbations of the 
fractional ionization $\delta x_e$.
Once we can get $\delta x_e$, we calculate $\alpha$,
which is defined by equation (\ref{alpha})
to estimate $\tau_{\rm R}$.  Using baryon density fluctuations $\deltab$
of the previous time step, we can calculate $\tau_{\rm R}$,
the growth rate and the growth factor.  
We can estimate $\deltab$  with
multiplying this growth factor by the amplitude of linear density
perturbations 
with the COBE normalization.

Strictly speaking, we may have to solve
radiative transfer during the recombination epoch to get precise 
value of fluctuations of  $\delta x_e$.
Instead, we have employed two approximation methods here. 

First method is to employ Saha's formula as Shaviv has done in his paper, 
which is a very simple but unrealistic assumption during and after 
the recombination epoch.  It is well known that even the fractional 
ionization 
itself leaves from the value by Saha's formula as recombination 
proceeds (see e.g., Jones \& Wise 1985). 
Therefore we next develop a perturbed treatment of the rate equation.
Detailed treatments of these two methods are described in Appendix B.
It is shown there that these two methods provide very different 
$\bar{x}_e$ and $\delta x_e$.

It is found that the nonlinear growth mechanism does not work at all 
for the fiducial CDM model
with $\Omega_0=1, \Omegab=0.04$, and $h=0.5$
if we employ the Saha equation for the evolution of  $\bar{x}_e$ and
$\delta x_e$.  It is because $\Delta V_0$ is too small during 
the recombination epoch, i.e., $1+z = 1500$ and $1100$.   
In this calculation, 
$1/\tau_{\rm H} + 1/\tau_{\rm D} > \tau_{\rm o} /\tau_{\rm R}^2$
for all time, which leads to the real part of $\omega$ of equation 
(\ref{solomega}) to be negative on small scales.

If we employ the perturbed rate equation to calculate $\delta x_e$,
however, the nonlinear growth mechanism is effective 
since the recombination process continues until much later epoch, 
$z \sim 800$ as is shown in Fig.~7 in Appendix B, when 
$\Delta V_0$ is large enough to induce the nonlinear growth 
by the drag force.
We obtain the nonlinear growth factor as a function of redshift $z$ 
in Fig.~5  for various COBE
normalized CDM models, i.e., the fiducial model with $\Omega_0=1$ and
$h=0.5$ and low density flat and open models with $\Omega_0=0.3$ and
$h=0.7$.  
For the purpose of comparison, different values of baryon
density, i.e., $\Omegab h^2=0.01$ and $\Omegab h^2=0.02$ are considered.
We adopt $3-\sigma$ peaks for the amplitude of the 
large scale velocity field.
The nonlinear growth mechanism 
works from $z=1100$ to $z=800\sim 600$ 
which depends on the value of $\Omegab h^2$.  

\begin{figure}
\centerline{\epsfxsize=15cm \epsffile{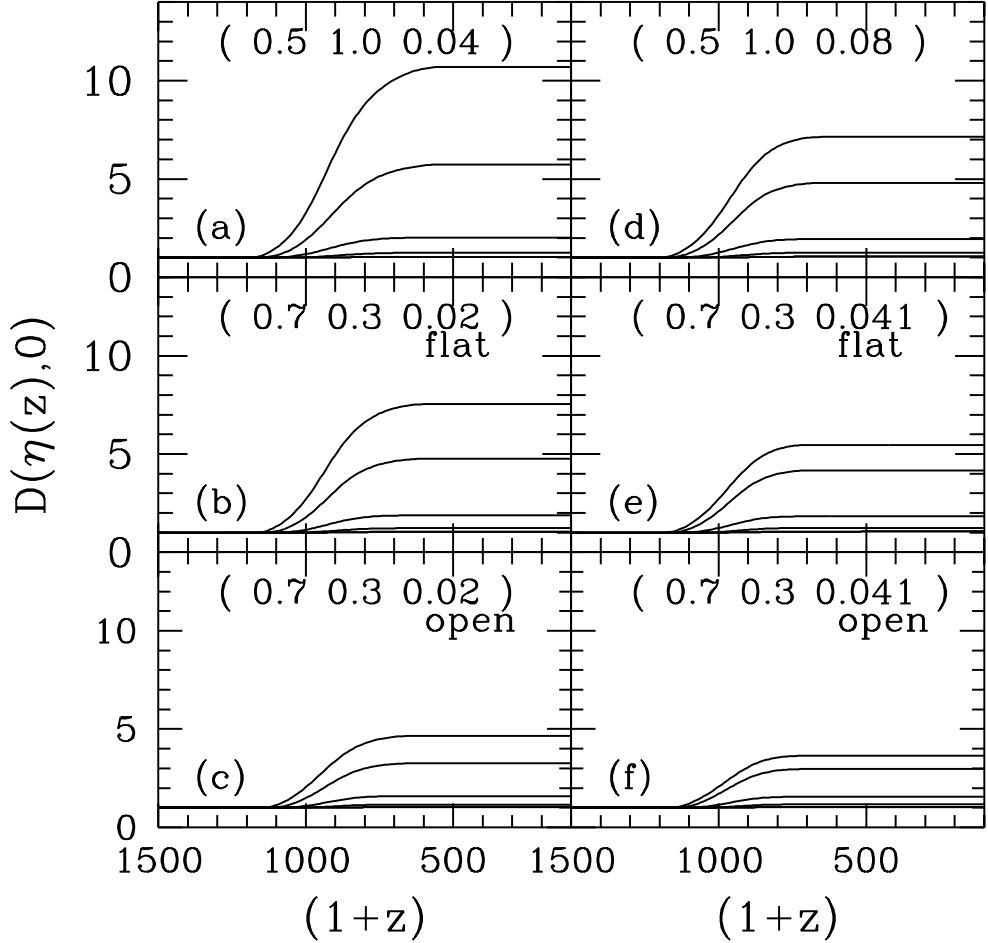}}
\caption{
 Growth factor $D(\eta (z),0) \equiv
 \exp\Bigl(\int_{0}^{\eta} d\eta' {\Re}[\omega(\eta')]\Bigr)$
 for various CDM models.
 The large scale velocity filed 
 $\Delta V_0$ is consistently solved by linear density perturbations 
 with COBE normalization (see Fig.~4). 
 The $3-\sigma$ velocity peaks ($\nu=3$) are 
 considered. The fractional ionization and its fluctuation are computed by 
 the rate equation. In parentheses, the cosmological parameters 
 $h,~\Omega_0$, and $\Omegab$ are shown. The mass scales of the
 baryon perturbations are $10^{-2},~10^{0},~10^{2},~10^{4}$, and $10^{6} 
 M_{\odot}$ from top to bottom, respectively. Panel (b) and (e) are low 
 density flat models while (c) and (f) are open models.
}

\label{Fig. 5}
\end{figure}

Fig.~6 plots the final growth factors 
as a function of wave number $k$.   We find that 
the growth factors of low density CDM models are less than 
the ones of $\Omega_0=1$ models.  
It is mostly because the COBE normalized rms velocity fields of low density
models are smaller when the mechanism effectively works ($z= 1100 \sim 800$) 
as is shown in Fig.~4.  
The strong $\Omegab$ dependence of the growth factor is also found in
Fig.~6.  The reason is
following.  The residual ionization at low redshifts scales as 
$\bar{x}_e \propto \Omega_0^{1/2}/(\Omegab h)$ (see e.g., Peebles 1993).
Therefore we have smaller $\bar{x}_e$ for larger $\Omegab$.  
The maximum growth rate can be written as (eq.~[\ref{max}]),
$\Re[\omega(\eta)]_{\rm max}=\left(
{\alpha \Delta V_0 / 2 c_s } -1
\right) /\left(2 \tau_{\rm D}\right)
.
$
\begin{figure}
\centerline{\epsfxsize=15cm \epsffile{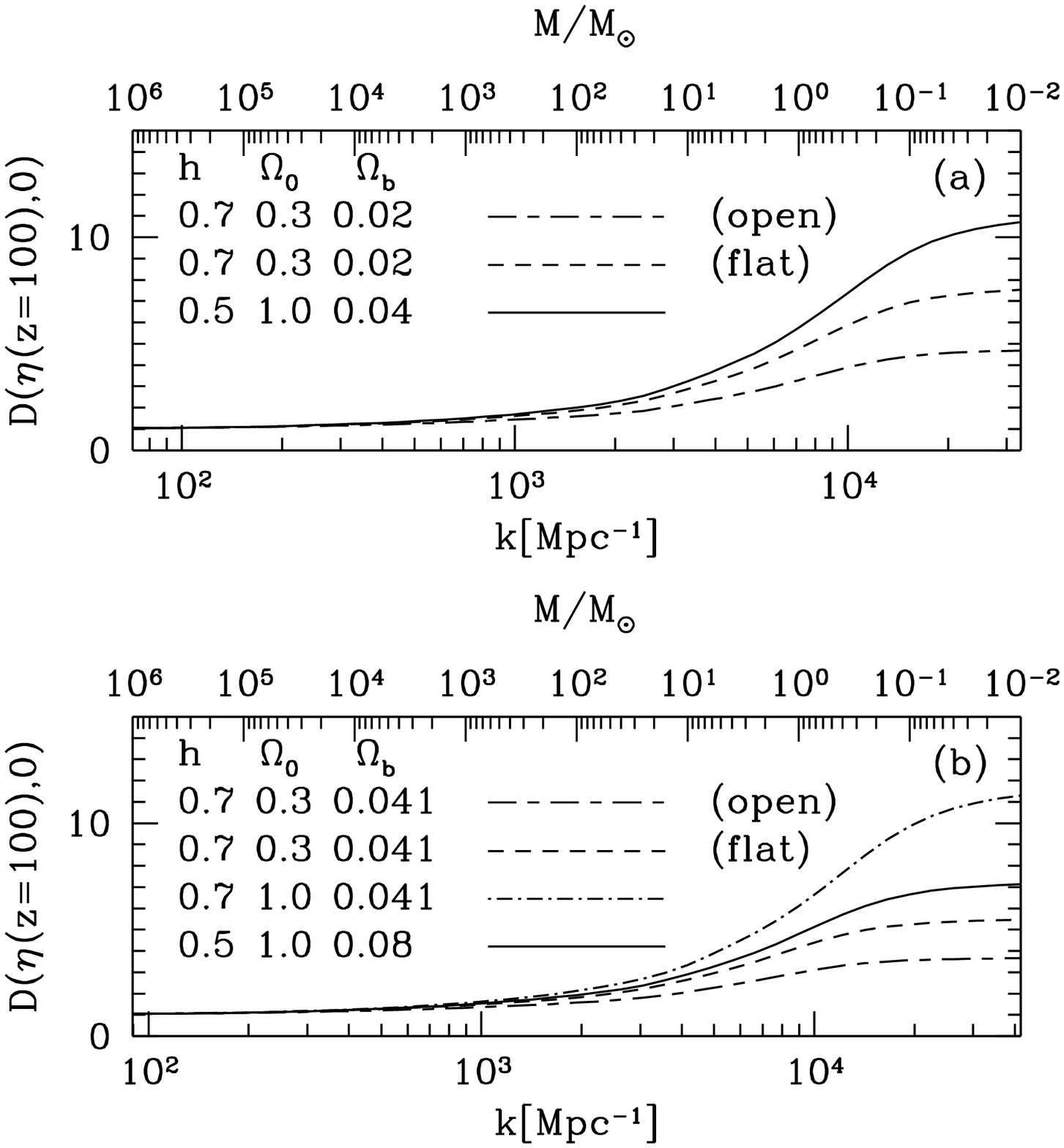}}
\caption{Final growth factor $D(\eta (z=100),0)$ for various cosmological 
models with $\nu = 3$ as a function of the wave number.
 The growth factor of the smallest baryon perturbation scale we chose
 ($M=10^{-2} \Msolar$) is
 close to the maximum growth factor $\exp\Bigl(\int d\eta 
\Re[\omega(\eta)]_{\rm max}\Bigr)$, where $\Re[\omega(\eta)]_{\rm max}$ is
defined in equation (\protect\ref{max}).
}

\label{Fig. 6}
\end{figure}
As is shown in Fig.~8 in Appendix B, the $\Omegab$ dependence of 
$\alpha$ is relatively weak because of cancellation in between 
$\bar{x}_e$ and $\delta x_e$.  The remaining dependence appears through 
$1/\tau_{\rm D} \propto \bar{x}_e$.  For larger $\Omegab$, therefore, 
$\bar{x}_e$ and the maximum
growth factor become smaller.

\section{Summary}

We investigate the evolution of the baryon density perturbations 
on very small scales, especially focus on the nonlinear growth 
induced by the shear velocity field on large scales, 
which is originally proposed by Shaviv (1998). 

First we study this nonlinear evolution 
with employing simple assumptions, i.e., the constant large
scale shear velocity field and the Saha equation for the calculations of
the fractional ionization and its fluctuation.  What we find is not
enormous amplification of density fluctuations but relatively mild
growth.  The reason of Shaviv's overestimation is because he has
neglected the diffusion
term which suppresses the growth.

It is found that the growth factor is very sensitive to the 
value of the large scale shear velocity field.  Only small difference 
($\Delta V_0 = 10^{-4}$ versus $6\times 10^{-5}$) causes 
huge divergence of the final growth factor (factor $50$ versus $2.5$).

Following the previous works (Paper I \& II), we next apply the nonlinear
growth mechanism to density perturbations of general adiabatic CDM
models.  In these works, it has been found that the baryon density 
perturbations are not completely erased by diffusion damping if there 
exists gravitational potential of CDM.  
They even grow before recombination under the balance between the 
radiative drag force and the gravitational force.  
We investigate the nonlinear growth of these baryon density
fluctuations with employing the perturbed rate equation which 
is proposed first time in this paper.  
The results are followings:
(1) The nonlinear growth is larger in smaller scales.  This mechanism
only affects the perturbations whose scales are smaller than
$10^2M_\odot$.
(2) The maximum growth factors of baryon density fluctuations 
by this mechanism for various COBE normalized CDM models are typically 
less than factor $10$ even if we take $3-\sigma$ large scale velocity
peaks.
(3) The strong $\Omegab$ dependence of the growth factor is found.  
This is because of the $\Omegab$ dependence of the residual ionization.
The fractional ionization $\bar{x}_e$, which is proportional to the inverse
diffusion time scale and the maximum growth rate,
is larger for smaller $\Omegab$.

How could this nonlinear growth mechanism affect on the structure
formation in the high redshift universe?  It is very interesting
coincidence that the typical scale of the nonlinear growth is about the
stellar size ($\lesssim 10^2M_\odot$).  However we need to carry out
high resolution numerical simulations to investigate the evolution of the
stellar scale density fluctuations on which a lot of other complicated 
physical processes, i.e., shock heating, UV radiation from first starts 
or AGN, cooling and so on, work.  Our quasi-linear perturbation analysis 
provides the initial condition for the calculations of 
such complicated structure formation in the 
high-z universe.   The transfer function of the baryon density
perturbations 
can be obtained by multiplying the nonlinear growth factor 
by the transfer function of the linear perturbations which is 
derived in Paper II.

\bigskip
\bigskip
\begin{center}
{\bf ACKNOWLEDGEMENT}
\end{center}
We thank Dr.~Tsuribe for stimulating discussions. We also thank
Professors Sato and Kojima for comments. The authors (K.Y. and H.N.)
are grateful to Y.~Kadoya for discussions related to the topics
of the present paper. 
This research was
supported by the Inamori Foundation, Sumitomo Foundation,
and in part by the
Grants-in-Aid program (11640280 \& 11640235) by the Ministry of Education,
Science, Sports and Culture of Japan. 
G.L. gratefully acknowledges the fellowship of Interchange Association.

\bigskip

\appendix

\section{Real Part of the Complex Square Root}

The derivation of 
the positive real part of the complex square 
root term in equation ({\ref{dispersion}}) is shown 
in this appendix. In general, the square root
of an arbitrary complex number can be rewritten as
\begin{equation}
\sqrt{A+Bi}=\sqrt{\sqrt{A^2+B^2}(\cos{\alpha}+i\sin{\alpha})}
=\sqrt{\sqrt{A^2+B^2}}\exp\left({i\alpha \over 2}\right)~ ,
\end{equation}
where $A$ and $B$ are real numbers and
\begin{equation}
\cos{\alpha}=\frac{A}{\sqrt{A^2+B^2}}\ \ , \
\sin{\alpha}=\frac{B}{\sqrt{A^2+B^2}}.
\end{equation}
Hence the real part of this square root is
\begin{eqnarray}
\Re[\sqrt{A+Bi}]&=&\sqrt{\sqrt{A^2+B^2}}\cos{\alpha/2}\nonumber \\
&=& \sqrt{\frac{1}{2}\left ( A +\sqrt{A^2+B^2} \right )}.
\end{eqnarray}
Let us now take
\begin{eqnarray}
A &=&\left ( \frac{1}{\tau_{\rm H}}+\frac{1}{\tau_{\rm D}} \right )^2
   +4\frac{1}{\tau_{J}^{2}}-4\frac{1}{\tau_{\rm o}^{2}}, \\
B &=& 4\frac{1}{\tau_{\rm R}^{2}}.
\end{eqnarray}
In the small scale limit, i.e., $k \rightarrow \infty$, we have 
$1/\tau_{\rm o}^{2} \gg 1/\tau_{\rm R}^{2} \gg 1/\tau_{\rm D}^{2}$, so that $|A| \gg B$.
We have to keep in mind that A is negative in this limit. Accordingly, 
we obtain the following approximation:
\begin{eqnarray}
\Re[\sqrt{A+Bi}] 
=\sqrt{{1 \over 2}\vert A \vert \left(-1+\sqrt{1+B^2/A^2}\right)}
&\simeq& \frac{B}{2\sqrt{|A|}} \simeq \frac{\tau_{\rm o}}
{\tau_{\rm R}^{2}} ~.
\end{eqnarray}

\section{Saha versus Rate Equations}

Some aspects of the baryon-electron system dealt with the recombination rate
have been discussed in the
previous papers (Paper I, II). In Paper I, the perturbed rate equation
for hydrogen has been presented, while the helium
fraction has been neglected and the time-evolution of the
equation has not been solved.
In the present paper we derive a new formula for the perturbation
of the fractional ionization taking the helium fraction into account.
The evolution of the fractional ionization $x_e$ 
is described as (Peebles, 1968)
\begin{equation}
-\frac{d}{dt}x_{e}=r_{e} n_{\rm b} \left (1-\frac{y_{p}}{2}\right )
\left \{x_{e}^{2}-\left[1-y_{p}-\left( 1-\frac{y_{p}}{2}\right)x_{e}
\right ]
\frac{x_{s}^{2}}{1-y_{p}-(1-y_{p}/2)x_{s}}\right\}C,
\label{peebles}
\end{equation}
where $r_e$ is the recombination coefficient, $x_s$ is the 
fractional ionization derived in use of the Saha equation
and $C$ is the suppression factor.
We employ the fitting formula for the recombination
coefficient by Pequignot et al. (1991) as 
\begin{equation}
r_{e} =10^{-13}\frac{aT_{4}^{b}}{1+cT_{4}^{d}}{\rm cm}^{3}{\rm s}^{-1},
\end{equation}
where the fitting constants are $a=4.309, b=-0.6166, c=0.6703$, and  
$d=0.5300$,
and $T_4$ is the baryon temperature in the unit of $10^{4}K$, i.e.,
$T_{4}\equiv T_{\rm b}/10^{4}K $.   
   From the Saha equation, $x_s$ can be derived as 
\begin{equation}
\frac{(1-y_p/2)^2x_{s}^{2}}{1-y_p-(1-y_p/2)x_s}=
\frac{(2\pi m_e T)^{\frac{3}{2}}}{n_{\rm b}(2\pi)^3}
e^{-{13.6 \rm eV}/{T}}.
\label{saha}
\end{equation}
The suppression factor $C$ in the right hand side is
\begin{equation}
C=\frac{1+K\Lambda n_{1s}}{1+K(\Lambda+\beta_{e})n_{1s}},
\end{equation}
where $n_{1s}$ is the number density of hydrogen in the electron
state, $\Lambda$ is the two-photon decay rate from the excited state to
the ground state, $\beta_{e}=r_{e}(m_{e}T/2\pi)^{3/2}e^{-3.4 {\rm eV}
/T}$
and $K=(a/\dot{a})\lambda_{\alpha}^{3}/8\pi$ with $\lambda_{\alpha}$
being the Lyman alpha photon wave length. 

Let us derive the perturbation equations on very small scales. We
introduce the perturbed variables,
\begin{eqnarray}
x_{e} & = & \bar x_{e}(\eta)+\delta x_{e}(\eta,\bf{x}) ,\\
x_{s} & = & \bar x_{s}(\eta)+\delta x_{s}(\eta,\bf{x}) ,\\
n_{\rm b} & = & \bar n_{\rm b}(\eta)(1+\delta_{\rm b}(\eta,\bf{x})), 
\end{eqnarray}
where $\bar{\null~\null}$ denotes the background quantities.
We ignore the temperature perturbation $\delta T$ since we are only 
interested in perturbations on very small scales where 
$\delta T$ is erased by Silk damping.

The unperturbed quantities $\bar x_{e}(\eta)$ and $\bar x_{s}(\eta)$
satisfy the same equations as (\ref{peebles}) and (\ref{saha})
with replaced $n_{\rm b}$, $x_{e}$, and $x_{s}$ by $\bar n_{\rm b}$, 
$\bar x_{e}$, and
$\bar x_{s}$, respectively. Then, from the Saha equation (\ref{saha}),
we find the linear perturbation equation for $\delta x_{s}$ as,
\begin{equation}
\delta x_{s}=-\frac{1-y_{p}-(1-y_{p}/2)\bar x_{s}}
  {2(1-y_{p})-(1-y_{p}/2)\bar x_{s}}\bar x_{s}\delta_{\rm b}.
\label{sahap}
\end{equation}
No perturbation of the fractional ionization can be generated
before recombination since $\bar{x}_s = (1-y_p)/(1-y_p/2)$.

   From equation (\ref{peebles}), on the other hand,
we can obtain the linear perturbation equation for $\delta x_{e}$ as,
\begin{eqnarray}
  -\frac{d}{dt}\delta x_{e}& = & C r_{e}\bar n_{\rm b}\left(1-y_{p}/2
  \right)\left \{ \delta_{\rm b}\bar x_{e}^{2}+\delta
x_{e}\left[2\bar x_{e}+
  \frac{(1-y_{p}/2)\bar x_{s}^{2}}{1-y_{p}-(1-y_{p}/2)\bar x_{s}}\right]
  \right. \nonumber \\
  & + & \left.\frac{ \delta C}{C}\left
[\bar x_{e}^{2}-[1-y_{p}-(1-y_{p}/2)
  \bar x_{e}]\frac{\bar x_{s}^{2}}{1-y_{p}-(1-y_{p}/2)\bar x_{s}}
  \right ] \right \} ,
\label{peeblesp}
\end{eqnarray}
where
\begin{equation}
\frac{\delta C}{C}=-\frac{K\beta_{e}\bar n_{\rm b}[(1-y_{p}-(1-y_{p}/2)
\bar x_{e})\delta_{\rm b}-(1-y_{p}/2)\delta x_{e}]}
{[1+K\Lambda \bar n_{\rm b}(1-y_{p}-(1-y_{p}/2)\bar x_{e})][1+K(\Lambda +
\beta_{e}) \bar n_{\rm b}(1-y_{p}-(1-y_{p}/2)\bar x_{e})]}.
\end{equation}

         Fig.~7(a) plots the fractional ionization $\bar x_s$ and 
$\bar x_e$ for the fiducial CDM 
model ($\Omega_0=1.0$ and $h=0.5$) and the $\Lambda$ CDM model
($\Omega_0=0.3$ and $h=0.7$) with high and low baryon densities, i.e.,
$\Omegab h^2 =0.01$ and $0.02$.

\begin{figure}
\centerline{\epsfxsize=15cm \epsffile{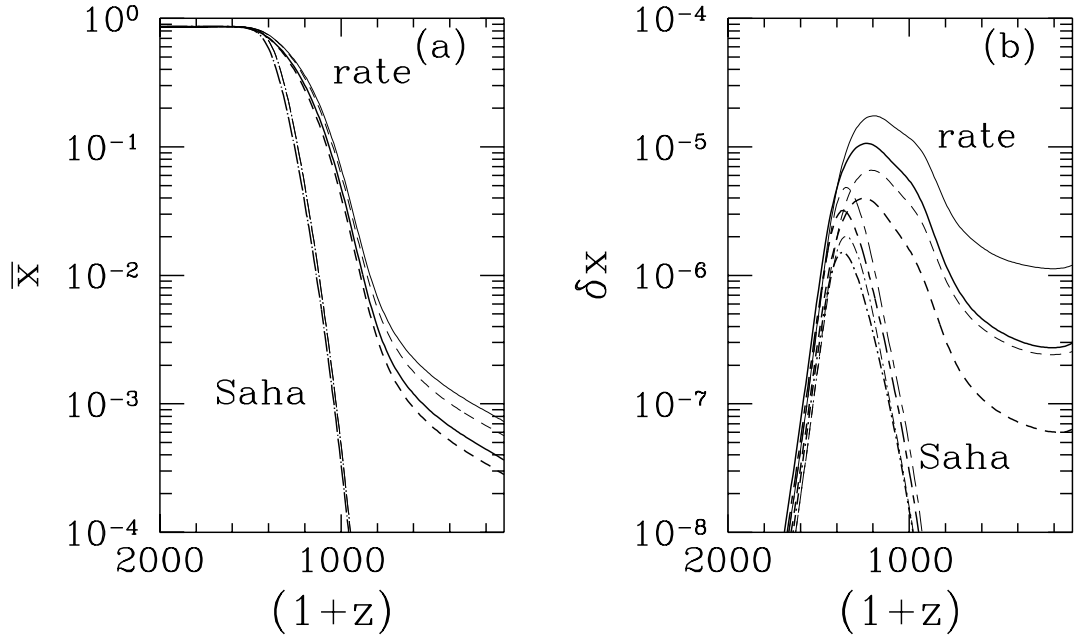}}
\caption{Fractional ionization (a) and its perturbation (b) from the 
Saha equation and the rate equation. 
We have two values of $\Omegab h^2$ ; $0.01$ and 
$0.02$ denoted by thin and thick lines, respectively. The solid
 and dashed lines describe the models with $\Omega_0=1.0$ and $h=0.5$ and 
$\Omega_0=0.3$ and $h=0.7$, respectively, for the rate
 equation. In panel (b), long dashed-short dashed lines denote the
 models with $\Omega_0=0.3$ and 
$h=0.7$ and dot-dashed lines denote $\Omega_0=1.0$ and $h=0.5$ for 
the Saha equation. 
Since the fractional ionization depends only on the value of $\Omegab h^2$
 in the Saha equation (\protect{\ref{saha}}), one can find only two
 lines for $\bar{x}$ with the Saha equation in panel (a). 
 The wave number $k$ is adopted as $7120 \rm Mpc^{-1}$.
}

\label{Fig. 7}
\end{figure}
The evolution of the Fourier coefficients of 
$\delta x_s$ and $\delta x_e$ are plotted in Fig.~7(b).  
Here we adopt the COBE normalized baryon density
perturbation with the wave number $k=7120 \rm Mpc^{-1}$ 
which corresponds to $1M_{\odot}$.   The nonlinear growth is 
taken into account to calculate the baryon density perturbation.
Before recombination, 
$\bar{x}_e \approx \bar{x}_s$ and $\delta x_e \approx \delta x_s$.  
As the recombination process proceeds, $\bar{x}_s$  rapidly decreases and
$\delta x_s$ falls as rapid as $\bar{x}_s$ because 
$\delta x_s$ is roughly proportional to $\bar{x}_s$ in equation (\ref{sahap}) 
if $\bar{x}_s \ll 1$. 
On the other hand, if we employ the rate equation, 
the recombination process proceeds much slower and eventually 
there remains the residual ionization as is shown in Fig.~7(a).  
This residual ionization results in the residual perturbation of 
the fractional ionization too.  Therefore thermal equilibrium is
not a good approximation for neither $\bar{x}$ nor $\delta x$. 

As is discussed in \S 2,  the drag force due
to the perturbations of the fractional ionization 
is parameterized by $\alpha / \tau_{\rm D}$.  
Here we define 
$\alpha_s \equiv \delta x_s/(\bar{x}_s\deltab)$ 
and $\alpha_e \equiv \delta x_e/(\bar{x}_e\deltab)$ 
to distinguish $\alpha$'s obtained 
by the Saha equation and the rate equation, respectively.

One can easily derive $\alpha_s$ from equation (\ref{sahap}) as 
\begin{equation}
\alpha_s=-\frac{(1-y_p)-(1-y_p /2)\bar x_s}{2(1-y_p)-(1-y_p /2)\bar x_s}.
\label{defalphas}
\end{equation}
Before the recombination epoch, $\alpha_s = 0$.  
And $\alpha_s=-1/2$ when the fractional ionization can be neglected. 
This value $-1/2$ is assumed in Shaviv's calculations.   
On the other hand, 
$\alpha_e$ is numerically calculated.  
Fig.~8 plots $\alpha_e$ and
$\alpha_s$ as a function of $(1+z)$ for various cosmological models. 
It is found that 
$\alpha_e$'s  are nearly
twice as large as $\alpha_s$'s during recombination 
when the nonlinear growth mechanism works.

\begin{figure}
\centerline{\epsfxsize=15cm \epsffile{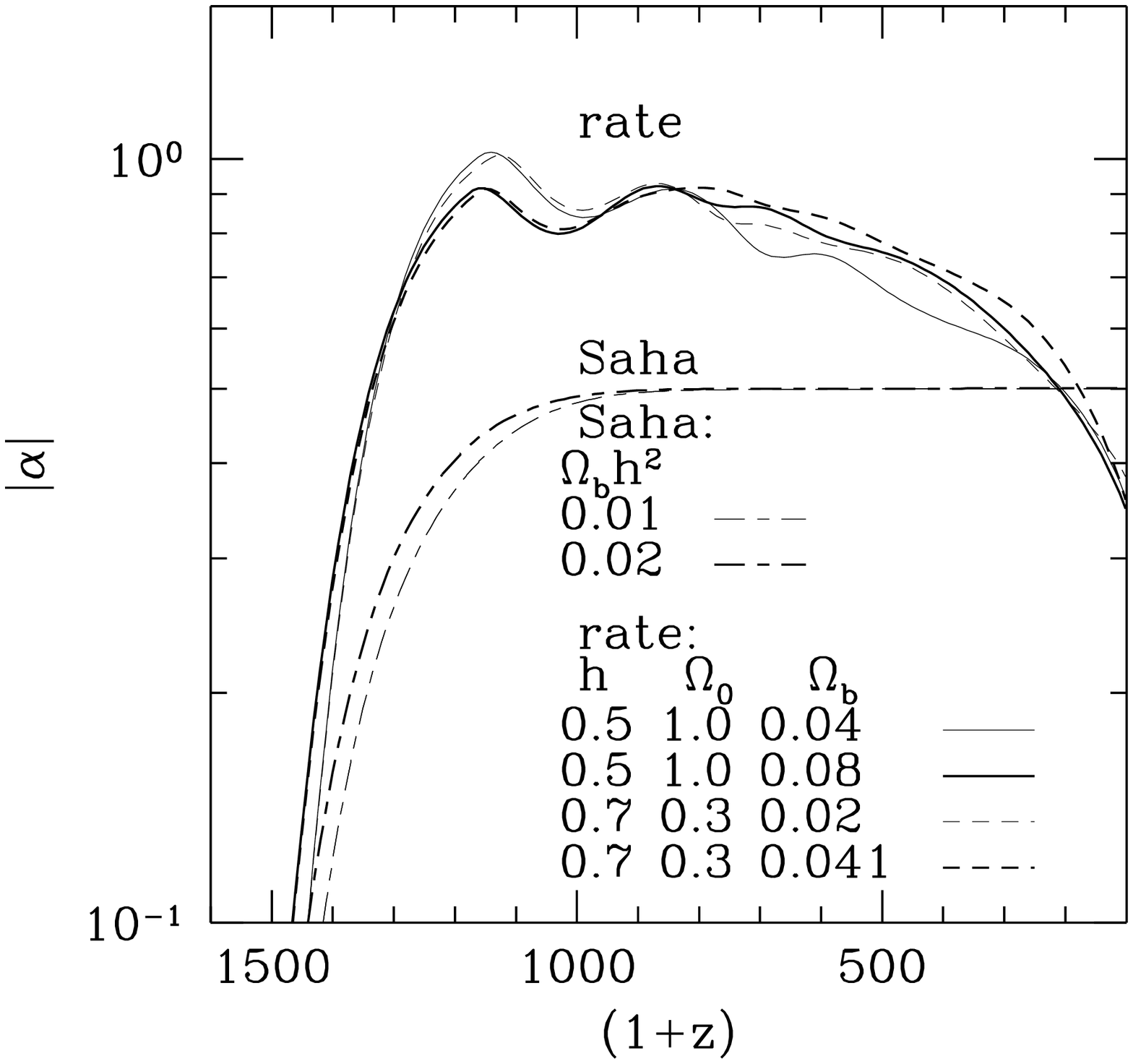}}
\caption{Parameter $|\alpha|$ (eq. [\protect\ref{alpha}]) from the rate equation and the Saha 
equation for various cosmological models. 
When the fractional ionization is small enough,
$|\alpha|=0.5$ in use of the Saha equation. The wave number $k$ is adopted 
as $7120  \rm Mpc^{-1}$.
}
\label{Fig. 8}

\end{figure} 
In Fig.~9, we plot $|\alpha|/\tau_{\rm D}$ for same cosmological models.  
Since the 
radiation drag force is proportional to $|\alpha|/\tau_{\rm D}$,
we expect larger amplification of the
baryon density perturbations if we employ the rate
equation instead of the Saha equation which Shaviv has used.  
Moreover, the nonlinear mechanism works until later epoch  ($z \sim 800$)
with the rate equation due to the residual $\bar{x}_e$ and $\delta x_e$.

\begin{figure}
\centerline{\epsfxsize=15cm \epsffile{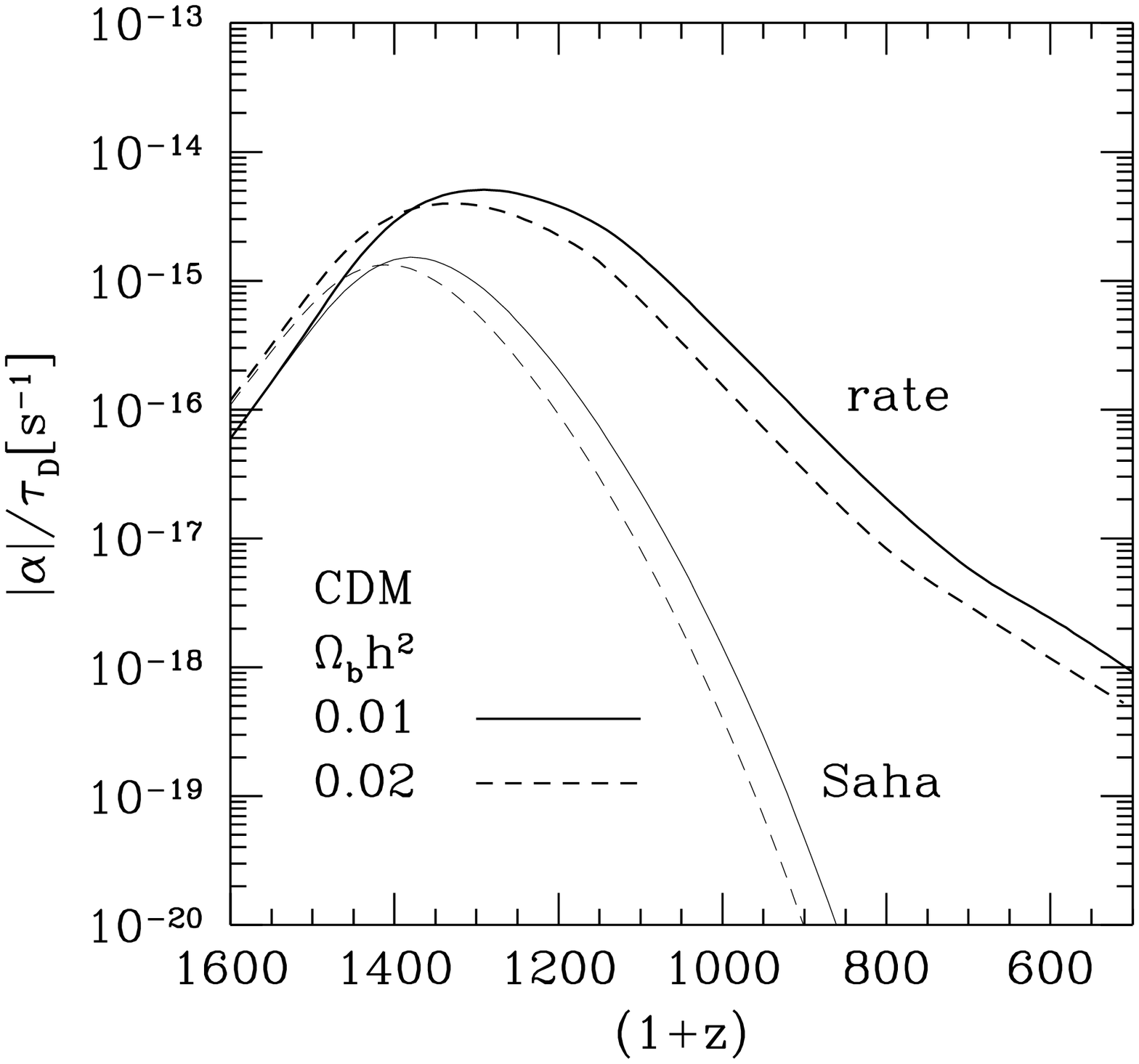}}
\caption{
Ratio of $|\alpha|$ to $\tau_{\rm D}$ from the rate equation (thick
 lines) and the Saha equation (thin lines) for the fiducial CDM models
 $\Omega_0 = 1.0,~\Omegab = 0.04$, and $h = 0.5$ (solid lines) and
 $\Omega_0 = 1.0,~\Omegab = 0.08$, and $h = 0.5$ (dashed lines).  The
 ratio depends only on $\Omegab h^2$ in use of the Saha equation (see
 eq. [\protect\ref{defalphas}] and eq. [\protect\ref{deftauD}]) but also
 depends weakly on $\Omega_0$ in use of the rate equation. The 
 quasi-nonlinear radiation drag force is proportional to
 $|\alpha|/\tau_{\rm D}$.  
} \label{Fig. 9}

\end{figure}


\end{document}